\documentclass[fleqn,usenatbib]{mnras}

\usepackage{mathptmx}

\usepackage[T1]{fontenc}

\DeclareRobustCommand{\VAN}[3]{#2}
\let\VANthebibliography\thebibliography
\def\thebibliography{\DeclareRobustCommand{\VAN}[3]{##3}\VANthebibliography}

\usepackage{graphicx}	
\usepackage{amsmath}	
\usepackage{amssymb}	
\usepackage{bm}
\usepackage{xcolor}

\title[Quantifying structure in SFRs]{Quantifying kinematic substructure in star-forming regions with statistical tests of spatial autocorrelation}

\author[B. Arnold et al.]{
Becky Arnold$^{1}$\thanks{E-mail: r.j.arnold@keele.ac.uk}
Nicholas J. Wright,$^{1}$
and Richard J. Parker$^{2}$\thanks{Royal Society Dorothy Hodgkin fellow}
\\
$^{1}$Lennard-Jones Laboratory, Keele University, Newcastle, ST5 5BB, UK\\
$^{2}$ Department of Physics and Astronomy, The University of Sheffield, Hicks Building, Hounsfield Road, Sheffield, S3 7RH, UK\\
}

\date{Accepted XXX. Received YYY; in original form ZZZ}

\pubyear{2022}

\begin{document}
\label{firstpage}
\pagerange{\pageref{firstpage}--\pageref{lastpage}}
\maketitle

\begin{abstract}
We investigate whether spatial-kinematic substructure in young star-forming regions can be quantified using Moran's $I$ statistic. Its presence in young star clusters would provide an indication that the system formed from initially substructured conditions, as expected by the hierarchical model of star cluster formation, even if the cluster were spatially smooth and centrally concentrated. Its absence, on the other hand, would be evidence that star clusters form monolithically. The Moran's $I$ statistic is applied to $N$-body simulations of star clusters with different primordial spatial-velocity structures, and its evolution over time is studied. It is found that this statistic can be used to reliably quantify spatial-kinematic substructure, and can be used to provide evidence as to whether the spatial-kinematic structure of regions with ages $\lesssim$ 6 Myr is best reproduced by the hierarchical or monolithic models of star formation. 
Moran's $I$ statistic is also able to conclusively say whether the data are $not$ consistent with initial conditions that lack kinematic substructure, such as the monolithic model, in regions with ages up to, and potentially beyond, 10 Myrs. This can therefore provide a kinematic signature of the star cluster formation process that is observable for many Myr after any initial spatial structure has been erased.

\end{abstract}




\section{Introduction} \label{introduction}

Young star-forming regions are immensely important for the study of a wide range of astrophysical phenomena. They are where the majority of stars form (\citealp{Elmegreen00, Lada03}; \citealp*{Krumholz19}), and by extension where the bulk of planets are born, and where many supernovae occur. Many of the fundamental processes that occur in these regions are not fully understood, and perhaps the most significant of these are the processes by which star clusters form.

It is debated whether star clusters form from mergers between smaller substructures according to the hierarchical model of star formation (\citealp{Efremov95}; \citealp*{Bonnell03}; \citealp{Gutermuth08, Henshaw14}; \citealp*{Vazquez-Semadeni17}; \citealp{Gouliermis18, Beattie19, Mondal21}), or whether they form monolithically (\citealp*{Kroupa01}; \citealp{Krumholz07, Banerjee14}; \citealp*{DaRio2014}). These two models offer very different views of the cluster formation process. They make different predictions about the local environment around young stars and planets, and the origins of unbound stellar systems such as OB associations \citep{Wright20}.

Critically, these two models make very different predictions about the initial spatial and kinematic structure of star-forming regions. Stars form from turbulent gas in giant molecular clouds, where the velocity dispersion scales with their size \citep{Larson81, Heyer04, Hennebelle12}. In the hierarchical model of star formation, stars form over a wide area and inherit the spatial and kinematic substructure from the gas they formed from \citep{Efremov98} before hierarchically collapsing under gravity to form a compact cluster. In the monolithic model of star cluster formation the gas collapses under gravity to form a single dense clump, in which star formation occurs. As a result, in the monolithic model initial spatial and kinematic structure is not expected.

The key difference between these models of star cluster formation is therefore whether star formation takes place before or after material is concentrated into the configuration of a dense cluster. In reality, star formation may be a combination or dilution of these hypothetical scenarios, but they represent useful models to explore and compare to observations.

Distinguishing between these models of star cluster formation has been difficult as forming star clusters are often obscured within molecular clouds. In the disc of the Milky Way, for example, searches for the progenitors of young massive clusters have so far failed to find giant molecular clouds with the necessary mass and density to form such systems monolithically \citep[e.g.,][]{Ginsburg12, Longmore14}. Furthermore, the availability of kinematic data (radial velocities and proper motions) to study the dynamics of recently formed clusters has, until very recently, been scarce.

Happily, the quantity and quality of kinematic data for stars in young star-forming regions and clusters has rapidly increased in recent years thanks to astrometric data from $Gaia$ \citep{Gaia21} and DANCe \citep{Bouy13}, and radial velocities from surveys such as APOGEE \citep{Majewski17} and the {\it Gaia}-ESO Survey \citep{Gilmore12}. Further, the availability of kinematic data is only likely to increase in the future thanks to upcoming surveys such as WEAVE \citep{Dalton14, Dalton20} and 4MOST \citep{deJong19}.

Kinematic data offers the prospect for identifying kinematic signatures of the star cluster formation process. For example, \citet{Wright19} recently discovered a trend of increasing velocity dispersion with stellar mass in the young cluster NGC~6530. They argued this could only occur if the cluster had formed hierarchically following the subvirial collapse of a substructured and extended distribution of stars \citep[see also][]{Parker16}, and might also indicate that the stars formed in a turbulence-dominated environment \citep{BonillaBarroso22}. This indirect kinematic signature offers the tantalizing prospect for directly testing models of star cluster formation using kinematics.

A possible avenue to pursue is to identify residual kinematic signatures left over from the formation of the star cluster. While spatial substructure is expected to be rapidly erased during mergers \citep{Binney08, Parker14}, any primordial kinematic substructure (as expected from the hierarchical model of cluster formation) is predicted to survive for longer \citep[e.g.,][]{Goodwin04}. 

Searching for residual kinematic substructure is a challenging task due to the high-dimensional and often messy nature of kinematic datasets, the short time period over which it might be observable, and the additional scatter in kinematic data introduced by binary and multiple stellar systems \citep{Gieles10, Cottar12}. The problem is further complicated by human biases which can lead us to see what we expect to see \citep{Nickerson98, Pannucci10, Rollwage20}, leading to potentially spurious identification or non-identification of structure.

To reliably identify residual kinematic substructure we need to utilize tried and tested statistical tools capable of quantifying the spatial autocorrelation of stellar kinematics. For this we turn to the field of spatial statistics, which have been utilized for many decades in the fields of climatology, environmental health, geochemistry, and ecology \citep[e.g.,][]{Mladenoff93, Law09, Wang10}. Such statistics are suitable for analysing data that are multivariate and spatially referenced, both attributes that are well-suited to the new generation of astrophysical data sets.

In this paper we explore one of the most commonly used spatial autocorrelation tests, Moran's $I$ statistic\footnote{The application of Geary's $C$ statistic \citep{Geary54} to this problem was also assessed and it was found to be ineffective. This is because Geary's $C$ statistic is better suited to identifying `hot spots' in spatial autocorrelation \citep{Ord95} rather than assessing a system as a whole.} \citep{Moran50}, which is primarily used in geosciences and ecology in order to study the spatial distribution of different phenomena, (e.g. \citealt{Anselin95, Fu14}; \citealt*{Liu15}; \citealt{Wagner15}; \citealt*{Fortin16}). It has also been used in astrophysics before to confirm the presence of structure in star forming regions \citep{Wright16}. Here we apply it to simulated star-forming regions that are generated to mimic the kinematics expected for different star-cluster formation scenarios. We demonstrate that this statistic can be used to differentiate between them, and that it can be used to detect these kinematic differences even after the simulated regions have evolved beyond the point that the initial spatial substructure has been erased. We also show that this method is robust against realistic observational biases.

\section{Methods}

In this section, we describe the $N$-body simulations used to simulate different cluster formation scenarios, as well as how Moran's $I$ statistic is calculated.

\subsection{\textit{N}-body simulations} \label{simulations}

To study the formation of star clusters, $N$-body simulations are used to evolve simulated star-forming regions forward in time from a variety of different initial conditions. The \textsc{\small{}kira} integrator in the \textsc{\small{}starlab} code \citep{PortegiesZwart99, PortegiesZwart01} is used to evolve the simulated regions to an age of 10~Myr. The level of kinematic substructure is measured every 0.1~Myr using Moran's $I$ statistic.

Six different sets of initial conditions are used in this work, composed of all combinations of two different sets of structural initial conditions and three different virial states. For each type of initial condition, 20 different realizations are generated and simulated so that stochastic variations in initial position and kinematics, and their resulting dynamics, can be explored. All the simulated regions are generated using a characteristic radius of 2~pc\footnote{For the fractal algorithm the characteristic radius is the radius of the spherical boundary imposed after the desired number of stars has been overproduced, see appendix \ref{frac_explanation} for details. For Plummer spheres the characteristic radius is twice the half-mass radius.}. The simulated star-forming regions in this paper all contain 1500 stars with masses drawn from a Maschberger initial mass function \citep{Maschberger13} with $\alpha = 2.3$, $\beta = 1.4$, and $\mu = 0.2$. A lower mass limit of 0.1 $M_{\odot}$ and an upper mass limit of 50 $M_{\odot}$ are imposed.

Two different types of structural initial conditions are generated. The first type are Plummer spheres \citep{Plummer11}, effectively smooth and spherical clusters without spatial or kinematic substructure that reflect the initial conditions expected by the monolithic model of star formation. They are generated by the method presented in \citet*{Aarseth1974}. This class of initial conditions is referred to as the unsubstructured class.

The other class of regions are generated such that they have kinematic structure, i.e. stars that are near each other have similar velocities, and a high degree of spatial substructure \citep[reflecting the hierarchical model of star formation,][]{Elmegreen96, Elmegreen04, Kruijssen12}. It is referred to as the substructured class, and the algorithm used to generate these regions is described in appendix \ref{frac_explanation}\footnote{The code for this algorithm can be found at \url{https://github.com/r-j-arnold/gen_fractal_star_clusters}.}.

Three different virial states are considered for the initial conditions, using the virial ratio $\alpha_{\rm{vir}}$. The virial ratio is the ratio of kinetic to potential energy in the system. A system with a virial ratio $<$ 0.5 will collapse on average (subvirial initial conditions), and one with a virial ratio $>$ 0.5 will tend to expand (supervirial initial conditions). A self-gravitating system in virial equilibrium  has $\alpha_{\rm{vir}} = 0.5$. In this paper $\alpha_{\rm{vir}}$ values of 0.3, 0.5 and 1.5 are used.

\begin{figure*}
	\includegraphics[width=\textwidth]{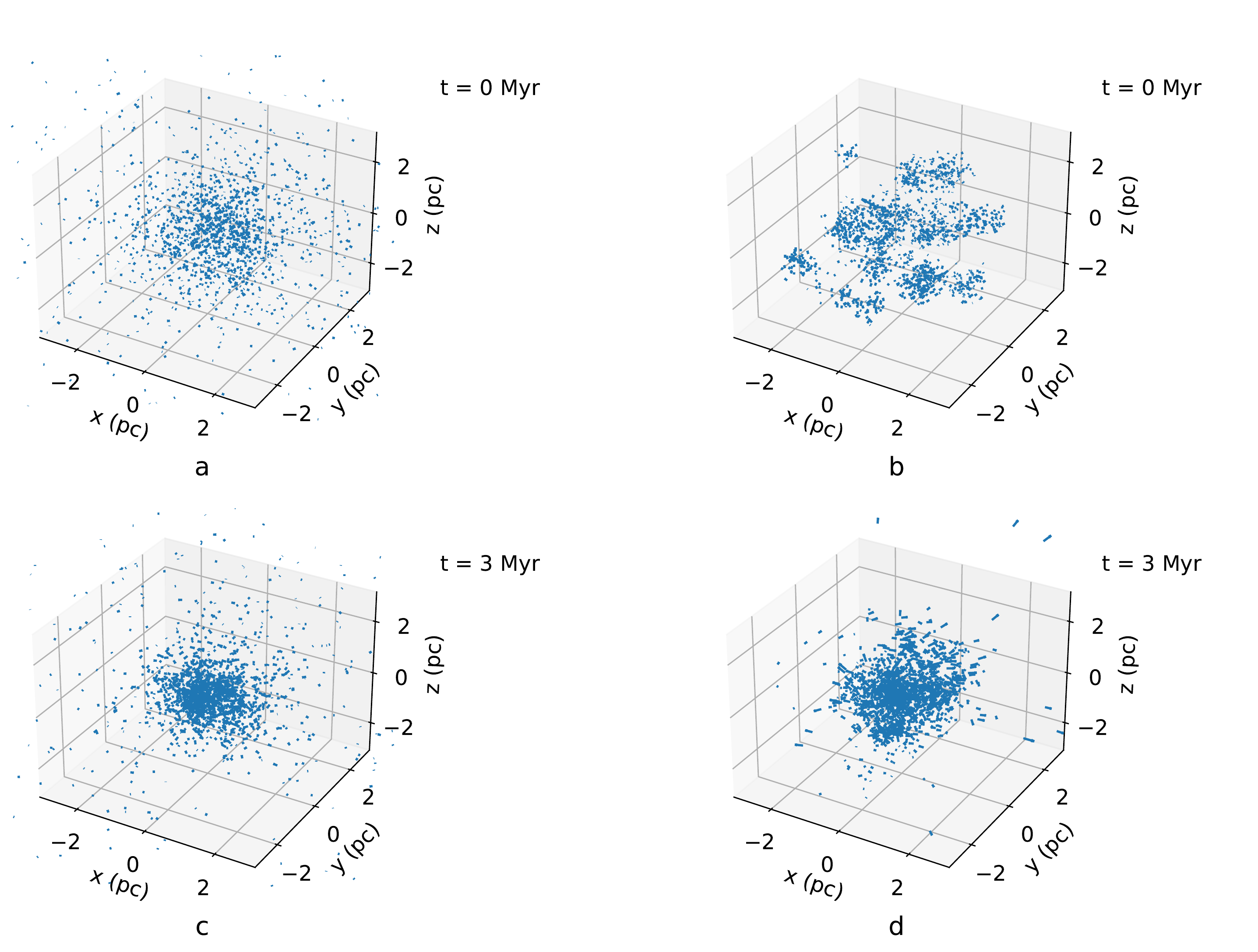}
    \caption{Simulated star-forming regions. The top row shows, from left to right, an unsubstructured region, and a substructured region. Both examples shown here are subvirial ($\alpha_{\rm{vir}} = 0.3$). These are shown at $t = 0$ in the top row and in the same order after they have been evolved to 3 Myr on the bottom row.
    }
    \label{fig_1}
\end{figure*}

Combining the structural initial conditions with the virial initial conditions gives a total of six different sets of initial conditions. Examples of initial conditions of each of the two structural classes are shown in the top row of Fig. \ref{fig_1}. Fig. \ref{fig_1}a contains an example of the unsubstructured class, and Fig. \ref{fig_1}b shows an example of the substructured class.

A key difficulty when studying structure in star forming regions is that obvious visual signatures of spatial substructure are erased within a few crossing times. To illustrate this the bottom row of Fig. \ref{fig_1} shows the two different sets of structural initial conditions evolved forward to an age of 3 Myr. It is readily apparent that after 3 Myr the initial spatial substructure (if there was any) has been erased, and it is very difficult to differentiate these evolved clusters visually as they are both centrally concentrated spheroids.

\subsection{Quantifying kinematic substructure} \label{quant_struct}

Moran's $I$ statistic for a given parameter $\mu$ is calculated as follows:

\begin{equation} \label{morans_i_eqn}
  I = \frac{N}{\sum_{i \ne j}{w_{ij}}}\frac{\sum_{i \ne j}{w_{ij}(\mu_i - \overline{\mu})(\mu_j - \overline{\mu})}}{\sum_{i}{(\mu_i - \overline{\mu})^2}}.
\end{equation}
 
\noindent where $w_{ij}$ is the weight given to a pair of data points, $i$ and $j$, in the calculation of Moran's $I$ statistic, and is calculated as
 
\begin{equation}
  w_{ij} = \frac{1}{d_{ij}}
\end{equation}

\noindent where $d_{ij}$ is the distance between points $i$ and $j$.

The expected value of Moran's $I$ statistic for data without any spatial autocorrelation is

\begin{equation} \label{expected_val}
  E(I) = \frac{-1}{N - 1}
\end{equation}

\noindent so in data without spatial autocorrelation (and where $N$ is large) Moran's $I$ statistic tends to zero.

If the calculated value of Moran's $I$ statistic for a data set is smaller than $E$($I$) this indicates spatial anticorrelation (i.e. similar values of $\mu$ tend to be far from each other). A Moran's $I$ value larger than $E$($I$) indicates spatial autocorrelation, i.e. that similar values of $\mu$ tend to be found close to one another. In the cases presented in this paper, $N$ is large enough that $E(I) \approx 0$ is a reasonable approximation in nearly all cases, so in general a positive Moran's $I$ value is taken to indicate spatial autocorrelation, and a negative Moran's $I$  value is taken to indicate spatial anticorrelation.

In this work, Moran's $I$ statistic is used to investigate the evolution of spatial autocorrelation of kinematic properties in star-forming regions. In calculating Moran's $I$ statistic according to equation \ref{morans_i_eqn} $N$ is the number of stars in the data set, and $d_{ij}$ is the distance between stars $i$ and $j$. Three kinematic quantities are examined (i.e. used as $\mu$); velocity in the $x$ direction, $v_{x}$, velocity in the $y$ direction, $v_{y}$, and stellar speed, $s$. The velocities are considered analogues to proper motions in RA and Dec, which are widely available thanks to {\it Gaia}. To reflect this, stellar speed is calculated in two dimensions, i.e. it represents the plane of sky speed, so

\begin{equation}
  s = \sqrt{v_{x}^{2} + v_{y}^{2}}.
\end{equation}

We use scalar quantities rather than vectors because Moran's $I$ statistic requires complex modification and more manual interpretation in order to be applied to vector data \citep[e.g.,][]{Liu15}. The choice of axis is arbitrary, so to mitigate its influence Moran's $I$ statistic for $v_{x}$ and $v_{y}$ are calculated separately and then the mean of the two is taken. This mean is referred to as Moran's $I$ statistic of $v_{2D}$, i.e. $I$($v_{2D}$). This quantity is less noisy than either $I$($v_x$) or $I$($v_y$) taken separately. Moran's $I$ statistic for $s$ is referred to as $I$($s$).

When calculating Moran's $I$ all stars at distances greater than two half-mass radii from the centre of mass of the cluster are excluded. This is because in reality not all original members of a star forming region are likely to be observed; stars that have been ejected from a region and have traveled a large distance away are significantly less likely to be positively identified as members by observers \citep{Schoettler20, Schoettler22}. Given the low likelihood that such outliers would be identified as members in observational data sets their exclusion from the calculation of Moran's $I$ statistic is justified.

It is important to understand that both global and local coherence in the parameter being studied can contribute to the calculated Moran's $I$ statistic. For example,
radial expansion of a star cluster, by definition, represents a global spatial autocorrelation in velocity direction (and so $v_{2D}$). This is because, in such a case each star's velocity vector is directly correlated with the star's position. This global velocity coherence contributes to increase the calculated $I$($v_{2D}$). In a similar vein, local velocity coherence, e.g. local groups of stars with similar velocities can increase $I$($v_{2D}$) even if the wider region the groups exist in is kinematically incoherent.

\section{Results} \label{results}

\begin{figure*}
	\includegraphics[width=\textwidth, totalheight=21cm, keepaspectratio]{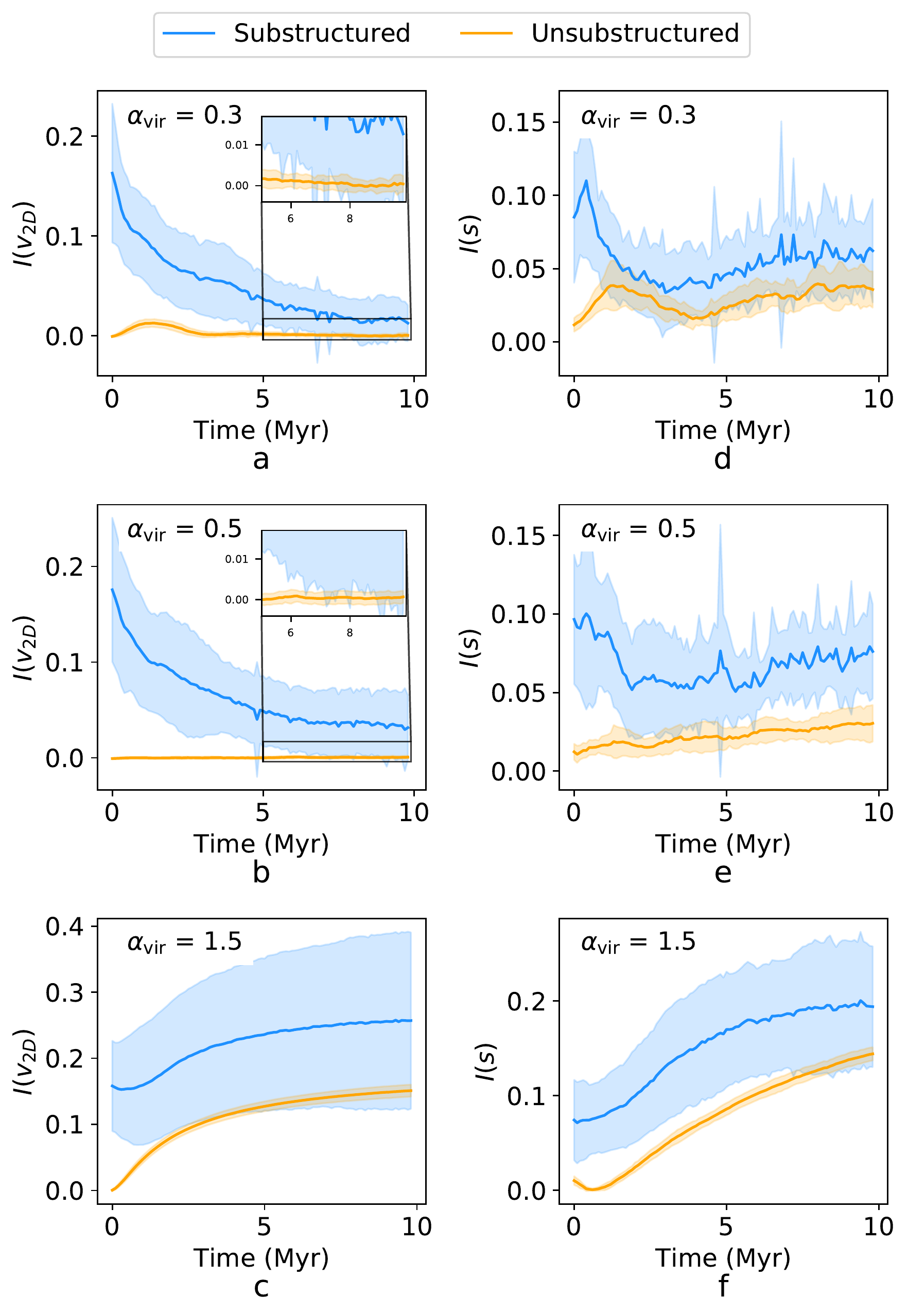}
    \caption{Moran's $I$ statistic versus time. The results for simulations with virial ratios of 0.3, 0.5, and 1.5 are shown in the top, middle and bottom rows, respectively. The $y$-axis of figures in the left-hand column is $I$($v_{2D}$), and in the right-hand column it is $I$($s$). In all subfigures the $x$-axis is time and the mean results for different simulation sets are shown in different colours, with their standard deviations depicted by the shaded areas. The results for substructured regions are shown in blue, and the results for the unsubstructured regions are shown in orange.}
    \label{fig_2}
\end{figure*}

The mean and standard deviation of $I$($v_{2D}$) and $I$($s$) at each time-step for each of the six sets of simulations is calculated, and the results are presented in Fig. \ref{fig_2}. The results for simulations with an initial virial ratio of 0.3 are shown in the top row of the figure, the results for simulations initially in virial equilibrium ($\alpha_{\rm{vir}} = 0.5$) are shown in the middle row, and those which are initially supervirial ($\alpha_{\rm{vir}} = 1.5$) are shown in the bottom row. 

The figures in the left-hand column show $I$($v_{2D}$) against time and the right-hand column shows $I$($s$) against time, where time is measured in Myr. For context the 10 Myr time-span of the simulations translates to between 2 and 4.25 crossing times for different initial virial ratios. This variation is due to the fact comparable regions with higher virial ratios have a higher velocity dispersion, and so shorter crossing times. Note that these are global crossing times, but in substructured regions the crossing time on the scale of individual stellar clumps, which itself will vary depending of the size and velocity dispersion of the clumps in question, is the more relevant metric in some contexts \citep{Allison09}. As a guide, 10 Myr translates to 25 - 50 crossing times on the scale of the smallest clumps in these simulations. 

In all subfigures the results for the substructured regions are shown in blue, and the results for the unsubstructured regions (i.e. Plummer spheres) are shown in orange. The standard deviation of the results for each simulation set is shown by the shaded areas of the corresponding colour. We emphasize that this standard deviation is not an uncertainty on Moran's $I$ statistic, which is calculated exactly from the simulated data according to equation \ref{morans_i_eqn}. Instead it is the $1 \sigma$ scatter in the results for regions that formed with the corresponding initial condition type.

\subsection{\texorpdfstring{The evolution of $I$($v_{2D}$) with time}{}} \label{v2d_results}

The evolution of $I$($v_{2D}$) against time is depicted by the panels in the left-hand column of Fig. \ref{fig_2}.

On average Moran's $I$ statistic is highest for the substructured regions, and it is lowest for the unsubstructured regions. This makes intuitive physical sense; the higher the degree of initial spatial-kinematic autocorrelation, the higher Moran's $I$ statistic is.

The evolution of $I$($v_{2D}$) over time is mainly determined by two physical processes:

\begin{itemize}
  \item The erasure of initial spatial-kinematic substructure (if generated with such) on both global and local scales as the regions become mixed and as stars undergo close gravitational interactions that randomize their velocities. This process has the impact of reducing $I$($v_{2D}$). 
  \item Fluctuations in the half-mass radii of the regions. As described in Section \ref{quant_struct}, global radial expansion causes $I$($v_{2D}$) to increase. Likewise global radial contraction represents a spatial-kinematic autocorrelation, and so also acts to increase $I$($v_{2D}$).  
\end{itemize}

The evolution of $I$($v_{2D}$) over time under different initial conditions will now be discussed in detail.

\subsubsection{Detailed discussion of the evolution of subvirial regions} \label{discussion_subvirial}

\begin{figure}
	\includegraphics[width=\columnwidth]{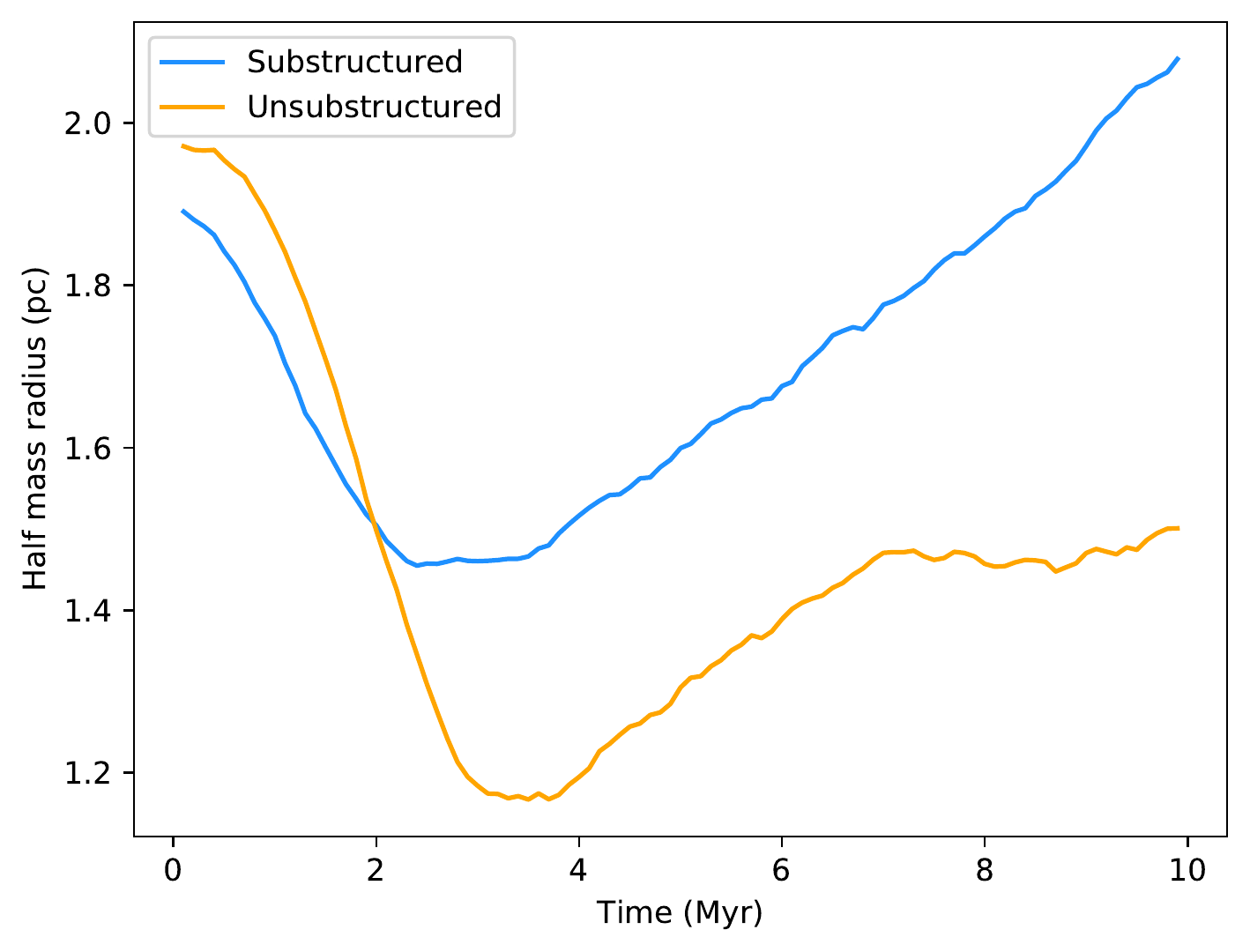}
    \caption{Half-mass radius plotted against time for simulations that are initially subvirial and with different types of initial substructure (see Fig. \ref{fig_2}a).}
    \label{fig_3}
\end{figure}

To aid this discussion the mean half-mass radii as a function of time for regions with subvirial initial conditions are presented in Fig. \ref{fig_3}.

The unsubstructured subvirial regions start off without any spatial autocorrelation of their velocities, so at 0 Myr Moran's $I$ statistic is approximately 0 (see Fig. \ref{fig_2}a). Because these regions are subvirial (i.e. collapsing), the half-mass radius of the regions rapidly shrinks (see the orange line in Fig. \ref{fig_3}). As previously discussed, this results in an increase in $I$($v_{2D}$), which peaks at $\sim$1.5 Myr i.e. when these regions are shrinking most rapidly (see Fig. \ref{fig_3}).    

Between 1.5 and 3.5 Myr the collapse of these regions slows down and stops (note the minimum in the half-mass radii of the unsubstructured regions in Fig. \ref{fig_3}), so $I$($v_{2D}$) returns to $\sim$ zero. The half-mass radii of these regions do steadily increase after this point due to the slow process of the many-body system dissolving over time, but this does not exhibit any spatially correlated kinematic signature and so doesn't cause $I$($v_{2D}$) to increase.

The substructured regions (blue lines in Fig. \ref{fig_2}a and Fig. \ref{fig_3}) start with $I$($v_{2D}$) $> 0$ at $t = 0$ Myr because they are generated with velocity substructure.
Like the previously discussed case of unsubstructured subvirial regions, these regions undergo rapid contraction (see Fig. \ref{fig_3}) which acts to increase $I$($v_{2D}$). However, this is overwhelmed by the reduction in $I$($v_{2D}$) induced by the destruction of the initial spatial-kinematic substructure by the dynamical evolution of the region (discussed in Section \ref{v2d_results}).

The ultimate impact of these two competing processes is $I$($v_{2D}$) drops steeply for the substructured regions, and then more slowly at later times ($>$ 5 Myr) once most of the dynamical evolution in the regions has occurred, and dynamical interactions remove any lingering kinematic substructure. The mean measured $I$($v_{2D}$) remains significantly above zero at times up to (and likely beyond 10 Myr).

\subsubsection{Detailed discussion of the evolution of initially virialized regions} \label{discussion_virial}

\begin{figure}
	\includegraphics[width=\columnwidth]{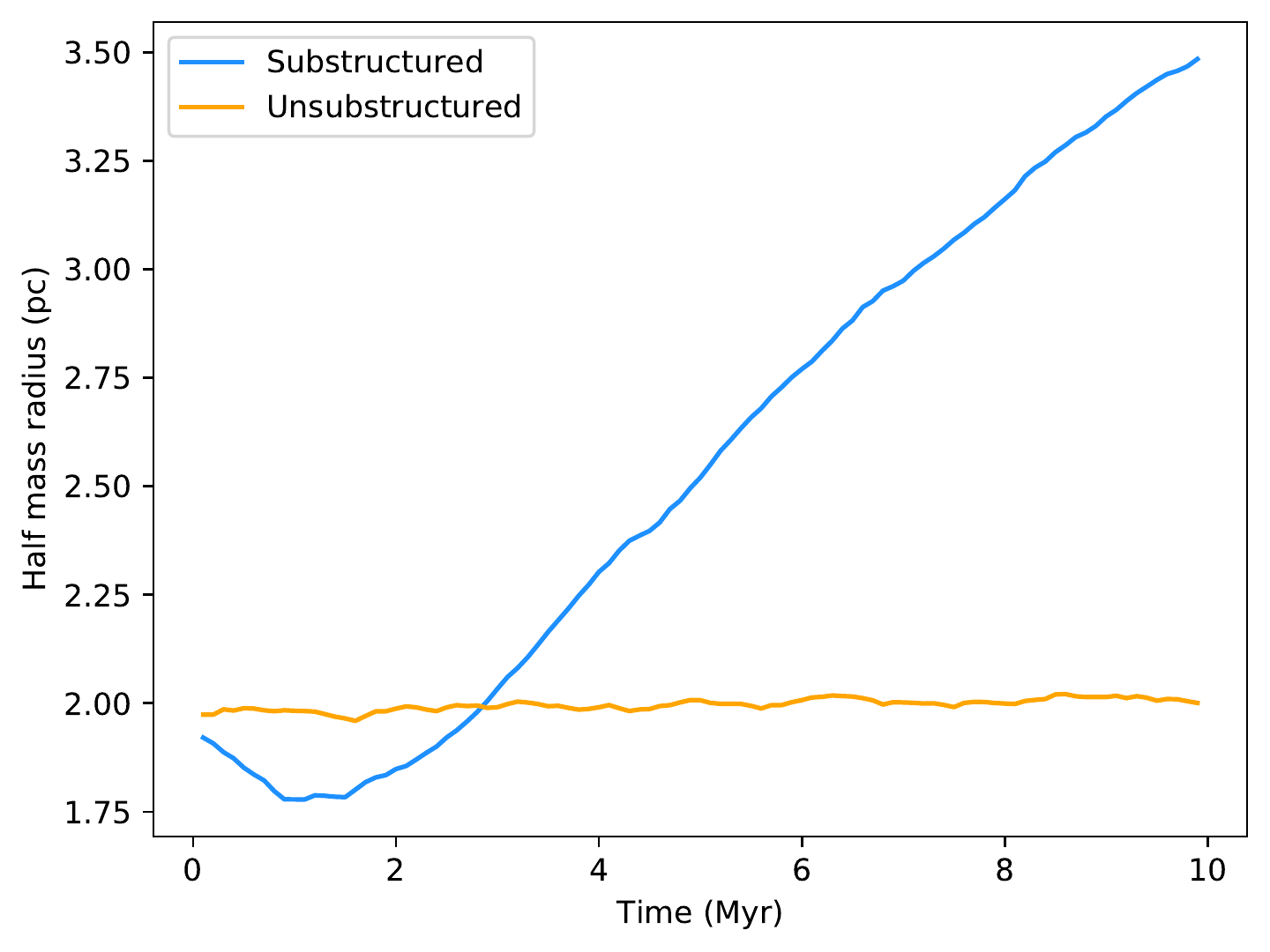}
    \caption{Half-mass radius plotted against time for simulations that are initially in virial equilibrium and with different types of initial substructure (see Fig. \ref{fig_2}b).}
    \label{fig_4}
\end{figure}

To aid this discussion the mean half-mass radii as a function of time for initially virialized regions are presented in Fig. \ref{fig_4}.

Fig. \ref{fig_2}b shows $I$($v_{2D}$) as a function of time for regions initially in virial equilibrium. Substructured, virialized regions follow the same basic trends as their subvirial counterparts (see Section \ref{discussion_subvirial}); Moran's $I$ statistic begins significantly above zero, and decreases over time as dynamical interactions erase the region's kinematic substructure. 

Despite being initialized in virial equilibrium the substructured regions do undergo collapse (see Fig. \ref{fig_4}), though to a much lesser degree than in the subvirial case. This can be seen by comparing their evolution in half-mass radii in Fig. \ref{fig_3} and Fig. \ref{fig_4}. These `collapses' are brief and the regions swiftly return to approximately their original half-mass radius, after which their half-mass radii continue to increase as the regions dissolve; it is more accurate to think of these `collapses' as kinetic and potential energy (which is already in virial equilibrium) being redistributed around the regions into a more stable arrangement rather than a collapse. Regardless, the impact of these pseudo collapses is small, and they are overpowered by the decline in $I$($v_{2D}$) induced by the dynamical evolution of the regions. As in the subvirial case, the mean $I$($v_{2D}$) remains significantly above zero up to and likely beyond 10 Myr.

Unsubstructured regions, on the other hand, in contrast to their subvirial counterparts remain at $I$($v_{2D}$) $= 0$. This is because they are already in equilibrium, and so do not undergo collapse (note their constant half-mass radius in Fig. \ref{fig_4}), and have no initial kinematic substructure for dynamical interactions to erase.

\subsubsection{Detailed discussion of the evolution of supervirial regions} \label{discussion_supervirial}

The overall trend of decreasing Moran's $I$ with time in Fig. \ref{fig_2}a and Fig. \ref{fig_2}b is reversed in the supervirial case, which is shown in Fig. \ref{fig_2}c. Here, $I$($v_{2D}$) increases in both cases, with the rate of increase reducing over time and plateauing. This is because the regions are supervirial and unbound so the regions expand, causing $I$($v_{2D}$) to increase.  

\begin{figure}
	\includegraphics[width=\columnwidth]{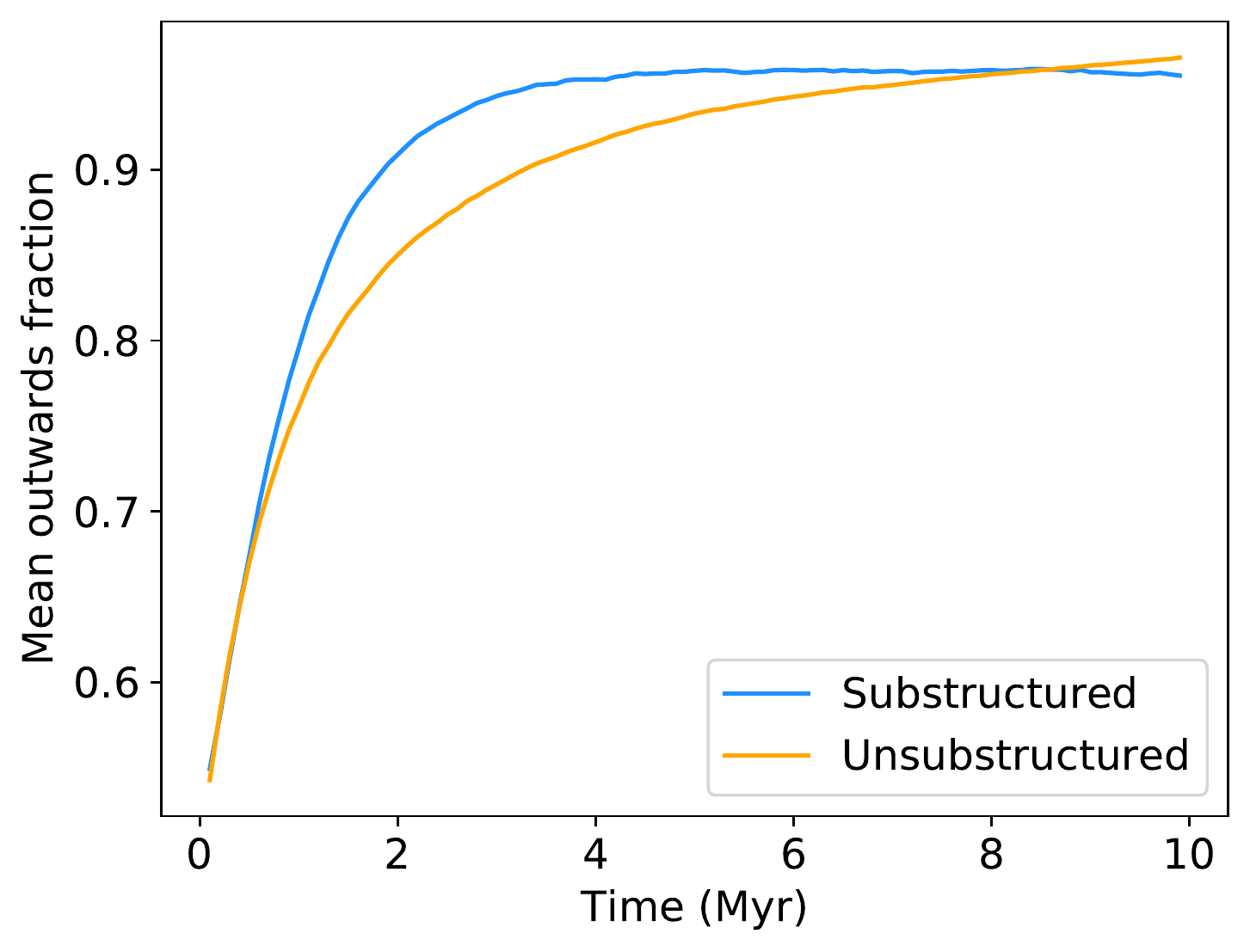}
    \caption{Plot of the mean fraction of stars moving away from the centre of mass in their star forming regions as a function of time for regions with supervirial initial conditions. The results for substructured regions are shown in blue, and the results for unsubstructured regions are shown in orange.}
    \label{fig_5}
\end{figure}

Fig. \ref{fig_5} depicts the mean fraction of stars moving away from their region's centre of mass as a function of time for the initially supervirial regions. Like $I$($v_{2D}$) in Fig. \ref{fig_2}c the fraction of stars moving outwards undergoes a rapid increase from $t = 0$ to $t \sim 2.5$ Myr, then plateaus. This further cements the explanation that radial motion due to expansion is the cause of the increase and plateau of $I$($v_{2D}$) observed in supervirial cases.

\subsection{\texorpdfstring{The evolution of $I$($s$) with time}{}} \label{vs_results}

Moran's $I$ statistic is applied to the speed, $s$, of the stars (right-hand panels of Fig. \ref{fig_2}). The results for $I$($s$) follow many of the same trends as the $I$($v_{2D}$) case:

\begin{itemize}
    \item Moran's $I$ statistic successfully identifies that there is initial kinematic substructure in $s$ for the substructured regions ($I$($s$) $> 0$ at $t = 0$ Myr for all blue lines). It likewise identifies the absence of initial kinematic substructure in the unsubstructured regions ($I$($s$) $\sim 0$ at $t = 0$ Myr for all orange lines).
    \item In the substructured subvirial and virialised cases $I$($s$) drops sharply at $t < 5$ Myr as the dynamical evolution of the regions erases substructure.
    \item The unsubstuctured, subvirial regions develop kinematic substructure as they undergo collapse, resulting in a peak in $I$($s$) at 1.5 Myr when the collapse is at its fastest.
    \item Both substructured and unsubstructured regions show a sharp increase in $I$($s$) followed by a plateau in the supervirial case as the regions expand.
\end{itemize}

Some differences are apparent in the evolution of $I$($s$) and $I$($v_{2D}$). Firstly, the results for $I$($s$) are noisier than those for $I$($v_{2D}$). Secondly, for subvirial and virialized regions in the $I$($v_{2D}$) case the trajectory of Moran's $I$ at late times is towards zero, but $I$($s$) at $t >$ 5 Myr shows a steady increase in all cases. This is because at later stages of dynamical evolution regions develop an active core and a slower moving halo of stars. This paradigm of fast moving core stars and slow moving halo stars represents a spatial correlation with $s$, so $I$($s$) increases. However, the directions of the stars (particularly in the core) are regularly randomized by gravitational interactions, so their is no directional correlation induced and $I$($v_{2D}$) tends to zero.

The impact of this is that in some cases $I$($s$) can separate substructured and unsubstructured initial conditions better at ages $\gtrsim$ 5 Myr than $I$($v_{2D}$). At earlier times it is also a useful metric to compute in addition to $I$($v_{2D}$) to support the presence or absence of kinematic substructure.

\section{Differentiating between models of star cluster formation} \label{diff_models}

The models presented in Section \ref{results} represent a variety of different initial conditions for the formation of star clusters and groups of stars. The theory of monolithic star cluster formation is best represented by the unsubstructured simulations (i.e., Plummer spheres) that form in virial equilibrium ($\alpha_{\rm{vir}} =$ 0.5). The $\alpha_{\rm{vir}} =$ 0.5 set is used because \textit{in situ} star formation (as predicted in the monolithic model) requires the region to remain largely static, so using initial conditions in virial equilibrium is appropriate. 

The hierarchical model of star formation, on the other hand, is best represented by the substructured simulations that are initialized in a subvirial ($\alpha_{\rm{vir}} =$ 0.3) state as the hierarchical model requires high initial substructure and multi-scale collapse that is not unphysically rapid. As shown in Fig. \ref{fig_1}, both of these initial conditions lead to the formation of a dense, centrally concentrated star cluster after a few Myr, as commonly observed in nature.

\begin{figure}
	\includegraphics[width=\columnwidth]{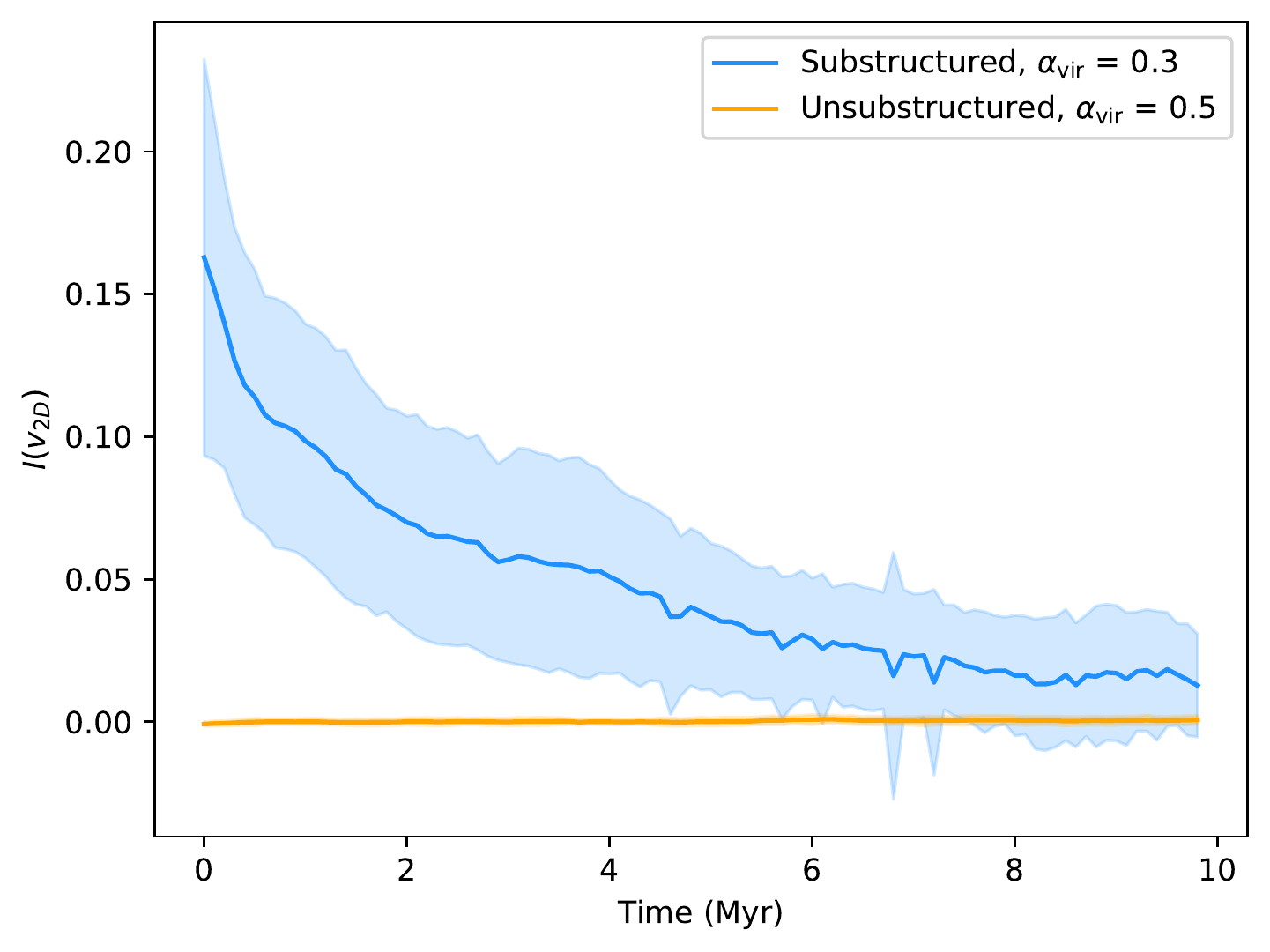}
    \caption{Plot of $I$($v_{2D}$) against time for two simulation sets. The hierarchical model of star formation analogue is plotted in blue, and the monolithic model analogue is plotted in orange. The standard deviation of the results for the simulation sets is shown by shaded areas of the corresponding colour. N.B. the standard deviation of the results for the monolithic model is very small meaning the shaded orange area is very thin, but it is present.}
    \label{fig_6}
\end{figure}

In Fig. \ref{fig_6} the mean $I$($v_{2D}$) of the simulations representing these two model analogues are plotted as a function of time, with the results for the hierarchical model analogue being shown in blue and the monolithic model analogue in orange. The standard deviation of the results of the simulation sets is shown by the shaded areas of the corresponding colours (note again that this does not represent the uncertainty on a given measurement, merely the range of observed values given the initial conditions). The scatter on the monolithic model results is extremely narrow; the results for this case are consistent with zero within a standard deviation $<$ 0.002 at all times. The mean $I$($v_{2D}$) is 0.00019, and the mean standard deviation is 0.0013. 

From this plot it is clear that Moran's $I$ statistic can be used to distinguish between these two initial condition types up to an age of 5--10 Myr depending on the precise evolution of the regions. Recall from Section \ref{simulations} that 5--10 Myr is a factor of two older than the time at which the models are virtually impossible to distinguish visually. By comparing the observed $I$($v_{2D}$) with the results presented here it can be determined whether the spatial autocorrelation of their stellar velocity vectors is consistent with the hierarchical model analogue presented here, the monolithic model analogue presented here, or neither.

It is worth noting that even beyond an age of $\sim$6 Myr (when the two shared areas begin to overlap in Figure \ref{fig_6}) Moran's $I$ statistic could be used to test the monolithic star cluster formation model. This is because the predicted values of Moran's $I$ statistic for the monolithic model analogue (shown in orange in Fig. \ref{fig_6}) occupy such a narrow band around zero that if a real star cluster had $I(v_{2D}) \gtrsim 0.002$ it could be confidently stated that the spatial autocorrelation of their stellar velocity vectors is inconsistent with the monolithic model analogue presented in this paper. There are different formulations of the monolithic model, but this result will most likely hold for any formulation which does not contain initial kinematic substructure, as the above result shows that Moran's $I$ can demonstrate the $absence$ of kinematic substructure to high precision.

\begin{figure}
	\includegraphics[width=\columnwidth]{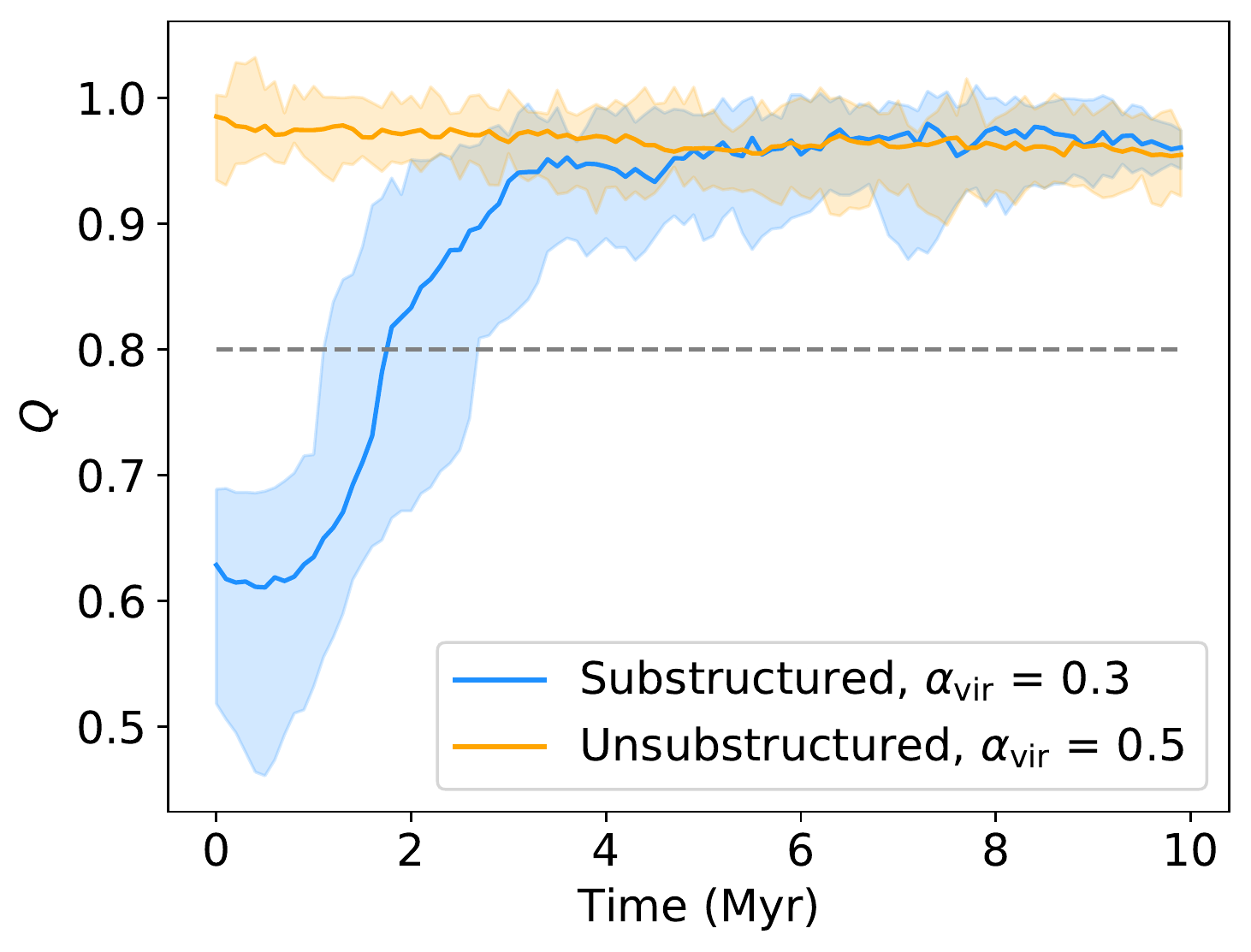}
    \caption{Plot of the $Q$ parameter \citep{Cartwright04} against time for two simulation sets. As in Fig. \ref{fig_6}, the hierarchical model of star formation analogue is plotted in blue, the monolithic model analogue is plotted in orange, and the standard deviation of the results for the simulation sets is shown by shaded areas of the corresponding colour.}
    \label{fig_7}
\end{figure}

As a comparison an alternative statistical metric, the $Q$ parameter \citep{Cartwright04}, is applied to these simulations. This widely used parameter is designed to quantify the degree of spatial substructure in star-forming regions (e.g. \citet{Rodriguez20, Nony21, Parker22}). Regions with a $Q$ parameter $<$ 0.8 are considered to be spatially substructured, and regions with a $Q$ parameter $>$ 0.8 are considered to be spatially smooth. In Fig. \ref{fig_7} $Q$ is plotted against time for the two model analogues presented in Fig. \ref{fig_6}. From contrasting these figures it is clear that the Moran's $I$ statistic is able to differentiate between the different models much longer than the $Q$ parameter. Moran's $I$ statistic also has the additional advantage over the $Q$ parameter that it can incorporate multidimensional spatial-kinematic data, whereas the $Q$ parameter is best suited to 2D data sets \citep{Cartwright09}.

\section{Observational uncertainties and biases} \label{obs_biases_section}

Observational data suffer from a number of uncertainties and biases that simulated data do not, and therefore it is necessary to test whether Moran's $I$ statistic still produces robust and reliable results when applied to kinematic data which have such imperfections. In order to do this we apply such biases and uncertainties to the simulated data and repeat the analysis to assess the impact this can have. The biases considered are incompleteness effects, contamination of the young star sample, and uncertainties on stellar velocities.

Binary stars also have the potential to introduce a number of observational uncertainties and biases. Close binaries with high orbital velocities will blur kinematic substructure when measured using radial velocities as the instantaneous velocity measurement will include a component from the orbital motion. Wide binaries could mimic very small-scale kinematic substructure as they will appear as a small, co-moving group of stars. Quantifying the impact of binaries is difficult as it will depend on a multitude of complex factors, such as the original binary fraction of the region being studied, the primordial semimajor axis distribution of those binaries, the exact dynamical history of the region, how far away the region is, and the properties of the observational setup used to measure stellar velocities.

However, a reasonable upper limit on the expected number of stars in observable binaries can be estimated. Assuming the use of \textit{Gaia} data (which can distinguish stars separated by more than 0.7 arcsec) and observing a cluster at a distance of 500 pc, binaries separated by more than 350 AU will be separately resolvable. 
Assuming binaries are randomly orientated, using the separation distribution of the companions of solar-type stars found by \citet{Raghavan10}, and presuming from their findings around 46 per cent of stars are in binaries (appropriate for solar-type stars, lower for less massive stars) we find around 12 per cent of stars are expected to be in resolvable binaries. This fraction is likely to decrease over time as wide binaries are destroyed via dynamical interactions \citep{ParkerMeyer14}. The impact of binaries on results will be further reduced by the fact that they only have heightened velocity coherence with their singular companion, but not with the other $N - 2$ stars in the region. Because of this the impact of their heightened local velocity coherence on the calculated Moran's $I$ statistic will be heavily diluted.

\subsection{Simulating observational biases}

To simulate the observational biases, we apply the following effects to the results of the $N$-body simulations before measuring Moran's $I$ statistic.

\subsubsection{Incompleteness} 

When a star-forming region is observed not all stars will be detected or identified to be young stars, most commonly because they are insufficiently bright to be detected. To mimic real incompleteness effects the lowest mass stars are removed from each data set (using stellar mass as a proxy for brightness) and the analysis is re-run. Note that this assumes that incompleteness is uniform across the observed region. Completeness fractions of 20, 40, 60, and 80 per cent are considered. As the simulated regions presented in this paper have stellar masses drawn from the Maschberger IMF \citep{Maschberger13} with the parameters outlined in Section \ref{simulations} a completeness of 20 per cent means stars of masses $\geqslant$ 0.63 $M_{\odot}$ are included in the analysis. For 40 per cent this is 0.33 $M_{\odot}$, for 60 per cent it is 0.20 $M_{\odot}$ and for 80 per cent completeness stars $\geqslant$ 0.14 $M_{\odot}$ are included.

\subsubsection{Contamination}

Background or foreground stars that are mistakenly thought to be part of a star-forming region can contaminate observational data sets. This observational bias is simulated by adding additional `observed' stars to the data set. The more similar a non-member star's spatial and kinematic properties are to those of true members the more likely it is that it will be included. To reflect this the positions of these contaminant stars are drawn from a Gaussian distribution with the same mean and standard deviation as the positions of the real members. The same approach is taken to draw velocities for the simulated contaminants. The impact on the results of contaminant fractions of 5, 10, and 15 per cent is examined.

\subsubsection{Kinematic measurement uncertainties}

Simulated measurement uncertainties of 0.25 km s$^{-1}$ and 0.5 km s$^{-1}$ are applied to the velocities by drawing `observed' velocities for each star from a Gaussian distribution with the mean of the true velocity of the simulated star and a standard deviation of the velocity uncertainty being mimicked. The uncertainties are applied to each velocity component separately. 

A velocity uncertainty of 0.25 km s$^{-1}$ corresponds approximately to a uncertainty of 0.1 mas yr$^{-1}$ at 500 pc. This value is chosen as it is the typical uncertainty on proper motions for $G =$ 18 magnitude sources in the $Gaia$ EDR3 catalogue \citep{Gaia21}. Observational uncertainties of 0.5 km s$^{-1}$ are also simulated in order to present a more extreme case, such as would be faced by fainter and/or more distant stars. The mean velocity dispersion of a single component of the velocities in the initial conditions of simulations presented in this paper is 0.66 km s$^{-1}$; the uncertainties applied here represent a significant fraction of a region's velocity dispersion.

\subsection{The impact of observational biases on the results} \label{impact_of_obs}

The observational biases described above are applied to the substructured subvirial simulations. This simulation type is chosen because it has a high degree of spatial autocorrelation in its kinematics, which makes it easier to observe the impact of observational biases on the measured Moran's $I$ statistic. 

\begin{figure*}
	\includegraphics[width=\textwidth, totalheight=22cm, keepaspectratio]{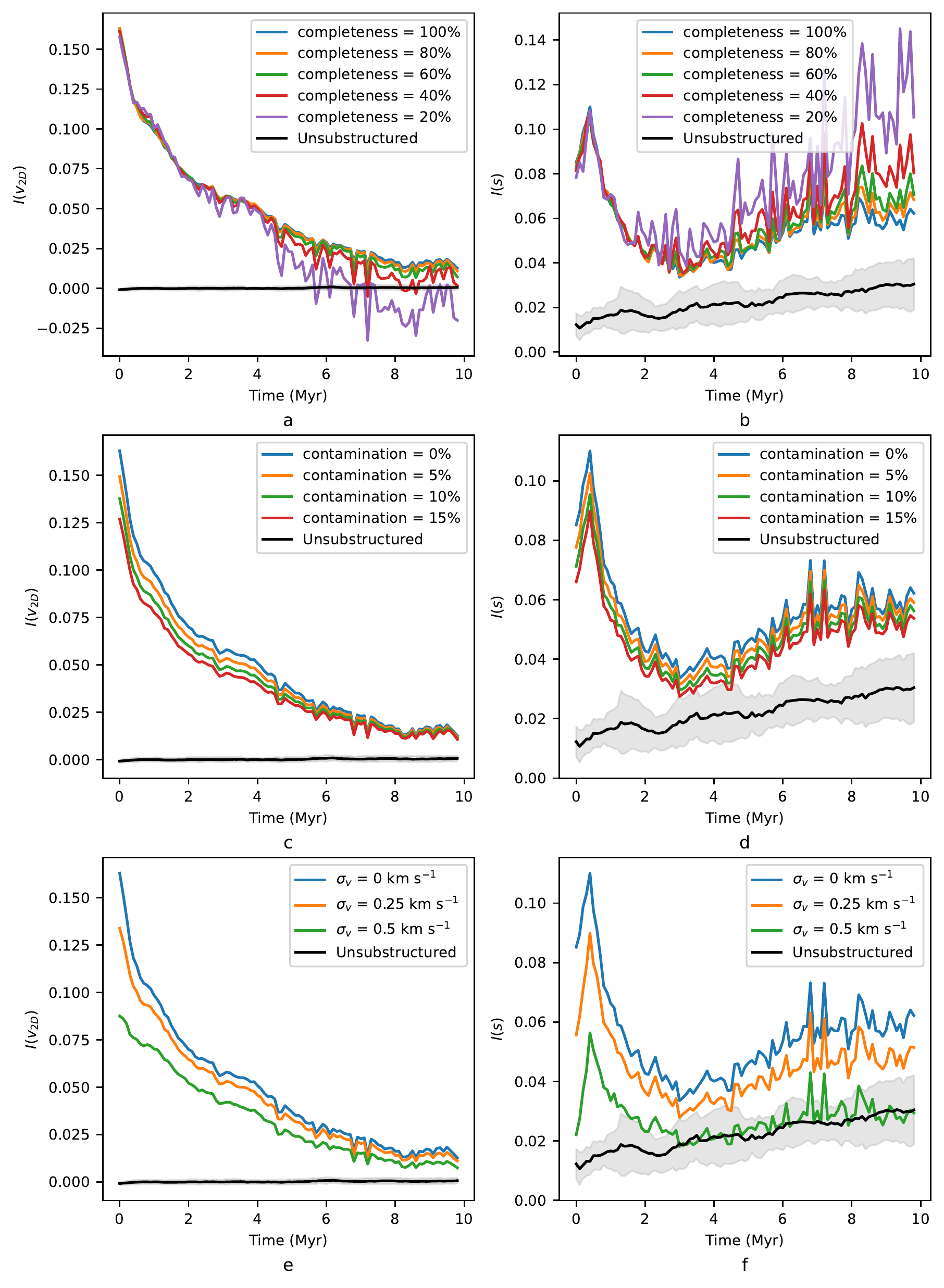}
    \caption{The results for substructured subvirial regions with different observational biases applied. The left-hand column shows the results for $I$($v_{2D}$) and on the right is $I$($s$). The impact of different completeness fractions is depicted on the top row, different contamination fractions are applied on the middle row, and on the bottom row the effect of measurement uncertainties on velocities is shown. The mean result for unsubstructured initially virialised regions is shown on all the subfigures in black, with the standard deviation of that simulation set shown in grey.}
    \label{fig_8}
\end{figure*}

Fig. \ref{fig_8} shows the calculated $I$($v_{2D}$) and $I$($s$) as a function of time after the above-described observational biases have been applied. The impact of incompleteness effects are depicted in the top row of the figure, contamination effects are shown on the middle row, and velocity measurement errors on the bottom row. As in Fig. \ref{fig_2} $I$($v_{2D}$) as a function of time is plotted in the left column and $I$($s$) as a function of time is shown on the right.

For reference, in all figures the mean result for unsubstructured regions initialized in virial equilibrium is depicted in black, and the standard deviation of that simulation type's results is shown by a grey-shaded area. This is done to illustrate where observational uncertainties would cause potential confusion between initial condition types (and by extension star formation models) and where it would not. 

In all cases the impact of observational biases on the results is to degrade them. In the case of reduced completeness one major consequence to the results is to increase the level of noise. This is logical because if there are fewer datapoints then increased noise on the measurement is to be expected. 

\begin{figure}
	\includegraphics[width=\columnwidth]{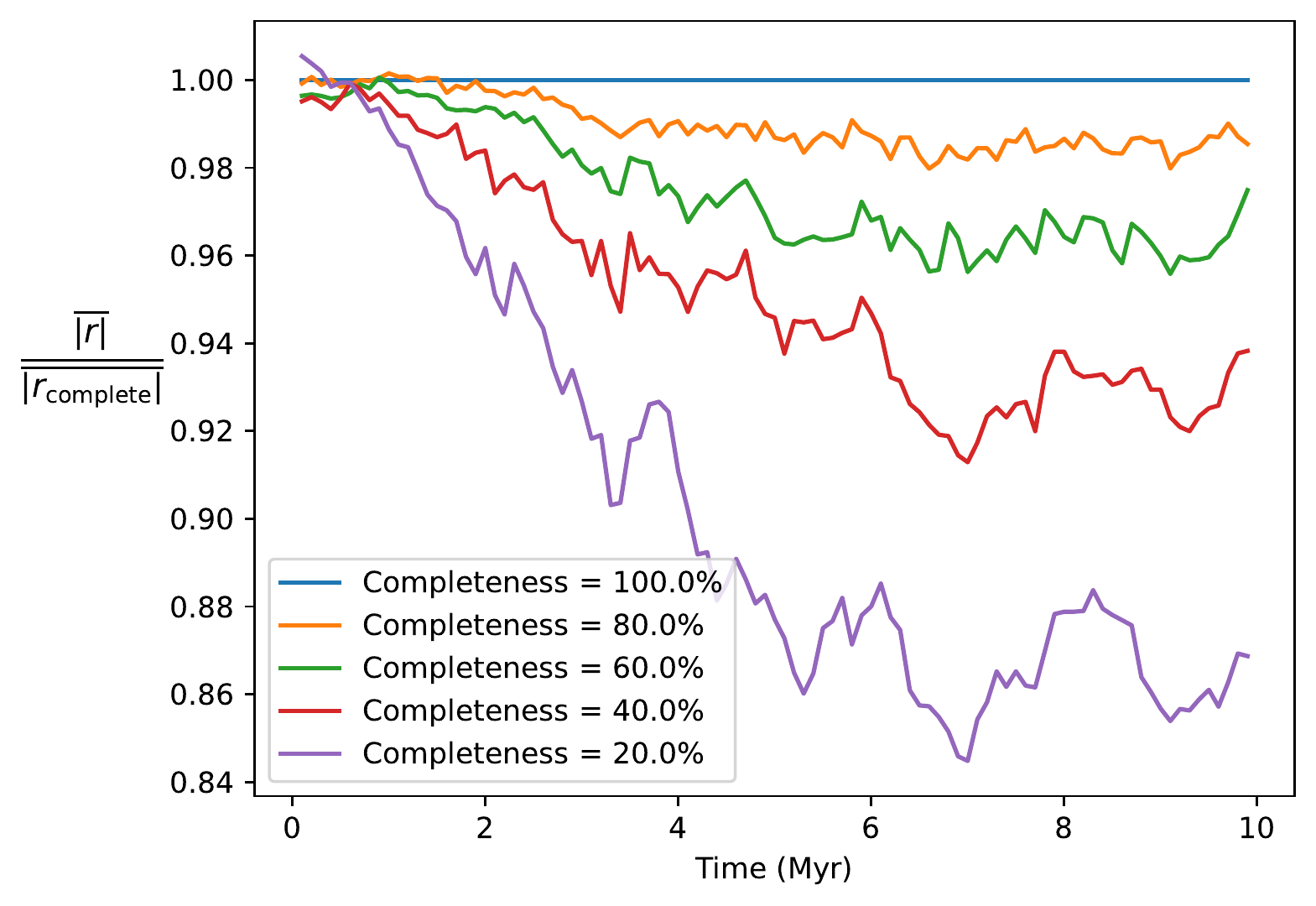}
    \caption{The mean distance of stars from their region's centre under different completeness cuts ($\overline{\lvert r\rvert}$) divided by the same for the complete sample ($\overline{\lvert r_{\rm{complete}}\rvert}$). This is plotted against time in Myr.}
    \label{fig_9}
\end{figure}

The other major impact of reduced completeness is to artificially lower the measured $I$($v_{2D}$), and increase the measured $I$($s$) at later times ($>$ 5 Myr). This is because after a few crossing times the regions become increasingly mass segregated. As a result, when completeness is low the sample becomes biased towards stars at small radii. To demonstrate this the mean distance of stars from the centre of the region $\overline{\lvert r\rvert}$ is calculated at different completeness cuts. These are divided by $\overline{\lvert r\rvert}$ for the complete sample, $\overline{\lvert r_{\rm{complete}}\rvert}$, and the evolution of this ratio over time is plotted in Fig. \ref{fig_9}.

The velocities of stars in the region's cores are regularly randomized by gravitational interactions, resulting in the reduction $I$($v_{2D}$) at low completeness. By comparing Fig. \ref{fig_8}a and Fig. \ref{fig_9} it is apparent that incompleteness causes the measured $I$($v_{2D}$) to deviate significantly from its complete value when $\frac{\overline{\lvert r\rvert}}{\overline{\lvert r_{\rm{complete}}\rvert}} \lesssim 0.94$. $I$($s$) is artificially increased at low completeness because stars in the halo have lower mean velocities than core stars; by disproportionately removing halo stars from the sample the gradient in $s$ with radius becomes more pronounced.

Despite this, Moran's $I$ statistic is able to trace the true signal well enough to differentiate these regions from unsubstructured regions (black line) at almost all completeness fractions and times. Overall the impact of even extremely low completeness on the results is relatively minor, indicating this method is robust against incompleteness.

Increased contamination (middle row) does not increase the degree of noise, instead it reduces the calculated value of Moran's $I$ statistic. Again this is logical as the spatial autocorrelation of a set of datapoints will be reduced if datapoints without any spatial autocorrelation are added, as they are when there is sample contamination. Again, the impact of this observational bias on the measured result is relatively minor and it is still possible to easily distinguish the substructured and unsubstructured regions from each other.

Finally, the impact of observational uncertainty is discussed. As with contamination the effect of velocity measurement errors is to dilute the measured spatial autocorrelation of the region's kinematics, and so lower the calculated Moran's $I$ statistic. $I$($v_{2D}$) is impacted much less severely than $I$($s$) because $v_{2D}$ is a measure of the change in velocity direction and the mean change in the direction of a velocity vector with a component drawn from a Gaussian centred on zero applied to it is zero. On the other hand, the mean magnitude of a value drawn from such a Gaussian (i.e. the change in speed, $s$) is $> 0$, so $s$ is the more strongly impacted quantity. 

Although the impact of observational bias is only presented here for one simulation type (substructured initially subvirial regions) the basic impact of biases on data with spatial-kinematic autocorrelation will be similar. That is to say incompleteness increases noise, and contamination and measurement uncertainties reduce the significance of the measured value of Moran's $I$ statistic. Nevertheless the impact of all but the most extreme observational biases on the results is relatively minor. 

Further, the only observational bias which can serve to cause an artificial increase in Moran's $I$ statistic is reduced completeness, and even then it has relatively little effect for completeness $\gtrsim$ 40 per cent, and almost no impact for regions less than 4 Myr old. As a result if a Moran's $I$ value is measured that is not consistent with $E$($I$), for a sample such that incompleteness could not be responsible then it can be determined that that signal is a real signature and not the result of observational bias.

\subsection{Impact of observational biases on model comparison} \label{obs_bias_model_comp}

Most observational data sets will suffer from a combination of the three types of observational bias discussed above and therefore we combine these effects and simulate the impact they have on the measurement of Moran's $I$ statistic. We perform this for three different levels of observational bias: mild, moderate, and severe. These three levels are outlined in detail in Table \ref{biases_table}, but as an example, in the mild case a completeness cut at 50 per cent is used, a contamination rate of 10 per cent, and a kinematic uncertainty of 0.25 km s$^{-1}$ is introduced. This is in addition to the standard cut of stars beyond two half-mass radii from the region's centres of mass, which is applied to all cases presented in this paper as discussed in Section \ref{quant_struct}.

\begin{table} 
\caption{The observational biases applied in mild, moderate, and severe cases.}
 \begin{tabular}{lccc} 
  \hline
  & Completeness & Contamination & Velocity error       \\
  & (per cent) & (per cent) &  (km s$^{-1}$) \\
  \hline
  Mild     & 50 & 10  & 0.25 \\
  Moderate & 30 & 20  & 0.5 \\
  Severe   & 10 & 30  & 1.0 \\
  \hline
 \end{tabular}
 \label{biases_table}
\end{table}

\begin{figure}
	\includegraphics[width=\columnwidth]{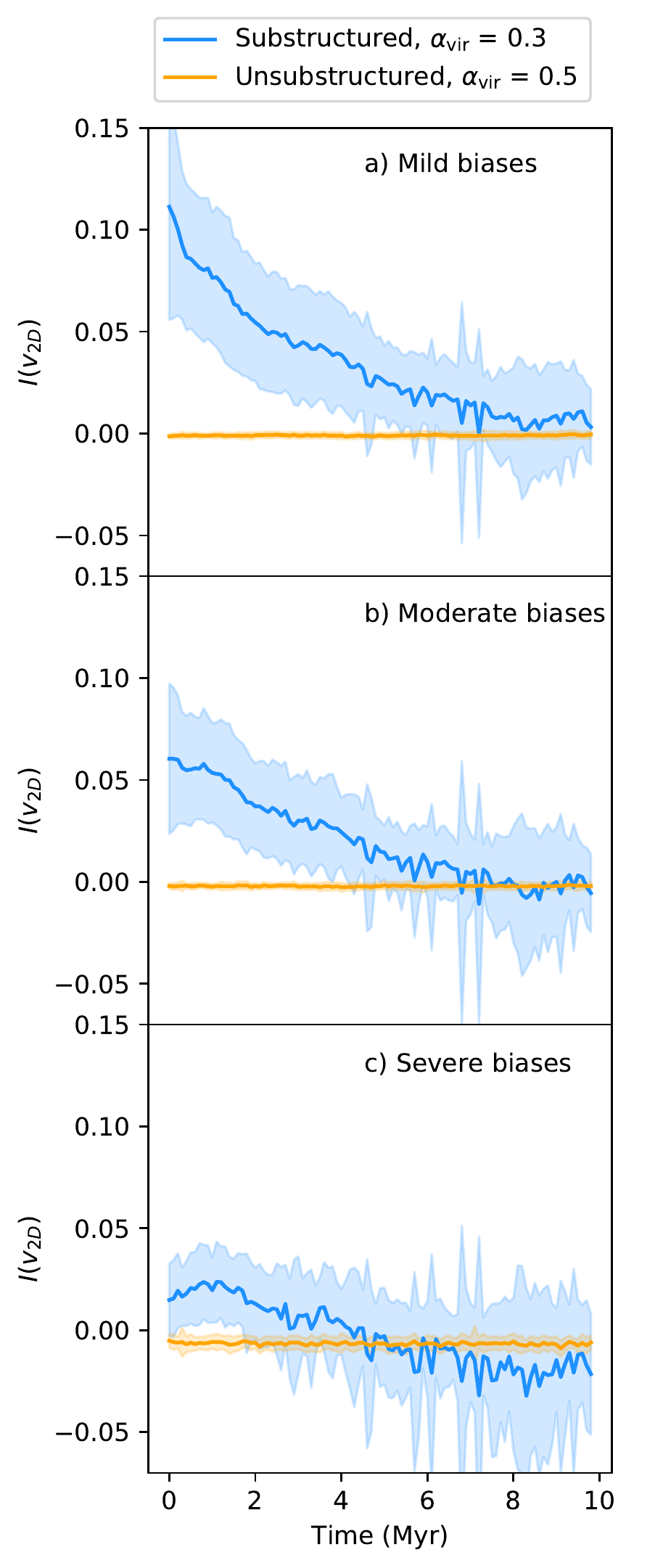}
    \caption{Plot of $I$($v_{2D}$) against time for two simulation sets that have had artificial observational biases applied. The hierarchical model of star formation analogue is plotted in blue, and the monolithic cluster formation model analogue is plotted in orange. The standard deviation of the results for the simulation sets is shown by shaded areas of the corresponding colour. The results are presented with mild (Fig. \ref{fig_10}a, top), moderate (Fig. \ref{fig_10}b, middle), and severe (Fig. \ref{fig_10}c, bottom) artificial observational biases applied. See Table \ref{biases_table} for further details.}
    \label{fig_10}
\end{figure}

The three levels of observational bias are then applied to the hierarchical collapse and monolithic cluster formation model analogues (discussed in Section \ref{diff_models}) and the results are presented in Fig. \ref{fig_10} as a function of time. The results for the mild case are shown in Fig. \ref{fig_10}a (top), the moderate biases results are shown in Fig. \ref{fig_10}b (middle), and the severe case is shown in Fig. \ref{fig_10}c (bottom).

By comparing Fig. \ref{fig_10} and Fig. \ref{fig_6} it is apparent that the effect of observational biases is to decrease the calculated $I$($v_{2D}$) and increase the noise on the results. Given the findings presented in Section \ref{impact_of_obs} this is entirely expected. However, the results in Fig. \ref{fig_10} still follow the same overall pattern as those with unbiased data, presented in Fig. \ref{fig_6}. The results for the hierarchical model analogue (shown in blue) $I$($v_{2D}$) still decrease with time, as in the unbiased case shown in Fig. \ref{fig_6}. 

In the case of the monolithic model analogue (shown in orange) the results remain close to zero at all times, but decrease a small amount in the more biased cases. This is because the expected value of Moran's $I$ statistic for data without spatial-kinematic autocorrelation, as is true for the monolithic case, is inversely proportional to $-N$ (recall equation \ref{expected_val} in Section \ref{quant_struct}). As completeness decreases, $N$ becomes smaller and therefore the expected value decreases too. The precise expected value varies between simulations and over time due to the exclusion of stars outside two half-mass radii of the centre of mass. However, even in the cases with the most severe biases the results for the monolithic case remain very close to zero. The calculated $I$($v_{2D}$) is -0.00095 $\pm$ 0.0017 if there are mild biases, -0.0021 $\pm$ 0.0019 if there are moderate biases, and -0.0066 $\pm$ 0.0036 in the case of severe biases. These are all decreased from the case without biases, where monolithic model simulations had $I$($v_{2D}$) of 0.00019 $\pm$ 0.0013. Nevertheless, it should be noted that the results for the monolithic model analogue show remarkably low scatter even with extreme observational biases applied. 

Critically, it is still possible to differentiate between these models using $I$($v_{2D}$) even when all the examined observational biases are applied if the biases are mild or moderate. Therefore in regions with such biases Moran's $I$ statistic may indicate whether the region formed with or without kinematic substructure, and therefore whether the hierarchical or monolithic model of star cluster formation best matches the observations. In a case with severe observational biases the degradation of the results is probably too extreme to do this. However, in all the cases presented here the results for the unsubstructured initial conditions occupy a very narrow band close to zero, with the precise expected value given by equation \ref{expected_val}. This is true at ages up to (and likely beyond) 10 Myr. Therefore if $I$($v_{2D}$) of real star-forming regions is not consistent with the expected value according to equation \ref{expected_val} given their $N$ within the tolerances found here (0.0017 given mild biases, 0.0019 given moderate biases, 0.0036 given severe biases) it can therefore be determined that observations are not consistent with having formed without substructure as would be expected in the monolithic model of star cluster formation.

\section{Summary}

We have investigated whether Moran's $I$ statistic can be used to quantify the level of kinematic substructure in star clusters and if it can be used to distinguish between the monolithic and hierarchical models of star cluster formation. The primordial morphology of star-forming regions is quickly erased during their initial dynamical evolution, as demonstrated in Fig. \ref{fig_1} where the two simulated regions with different initial conditions, shown on the top row, are virtually indistinguishable by-eye after evolving for just 3 Myr, as shown on the bottom row. We apply Moran's $I$ statistic to simulated star-forming regions with different initial conditions and find that it can successfully quantify their kinematic structure, and can differentiate between them (see Fig. \ref{fig_2}).

The evolution of Moran's $I$ statistic for simulated analogues of the monolithic and hierarchical star cluster formation models are closely examined. From Fig. \ref{fig_6} we find that Moran's $I$ statistic can differentiate between them at ages $\lesssim$ 6 Myr. Further, it may be used to provide strong evidence that regions formed with some initial kinematic substructure (against the expectations of the monolithic model of star formation) in regions even older than this. This is primarily because our simulated unsubstructured, initially virialized regions, which represent clusters that formed monolithically, show very low values of $I$($v_{2D}$) at all times (a magnitude smaller than 0.01 even in our worst-case scenarios). Therefore, if $I$($v_{2D}$) were to be measured in a real star cluster and found to be significantly outside of this range  then this would indicate a region formed with some degree of spatial and/or velocity substructure. A reasonably accurate age estimate of a given region will be of great help in effectively exploiting this metric. 

It is difficult to distinguish between regions that formed in a subvirial state or in virial equilibrium. While it is true that regions initialized with lower virial ratios tend to have slightly lower $I$($v_{2D}$) at later times the fact remains that Fig. \ref{fig_2}a is very similar to Fig. \ref{fig_2}b, and Fig. \ref{fig_2}d is very similar to Fig. \ref{fig_2}e. However, there is a very clear difference between the two bound cases and the unbound case presented in the bottom row of Fig. \ref{fig_2}. From this it is clear that if $I$($v_{2D}$) is larger than 0.1 in a region older than around 5 Myr then the region most likely formed in an unbound state.

The impact of observational biases on the results is investigated. In summary, the impact of incompleteness is to increase the degree of noise on the detected signal, and in the case of extreme incompleteness ($\leq$ 40 per cent of stars detected) artificially decrease $I$($v_{2D}$) and increase $I$($s$). The impact of stellar contaminants and measurement uncertainties on stellar velocities is to decrease the calculated Moran's $I$. These effects should be kept in mind when utilizing Moran's $I$ statistic to investigate the degree of kinematic substructure in a region and the implications for its origins, in order to evaluate results with proper context. Overall these investigations find that the impact of observational biases on Moran's $I$ statistic is minor in all but the most severe cases. It can still be used up to ages of 6 Myr to determine from observational data whether a region formed with kinematic substructure consistent with the hierarchical model, an absence of kinematic substructure (as would be expected by the monolithic model), or a degree of substructure which is consistent with neither model.

At greater ages, and in the case of severe observational biases this is not possible as the results for the monolithic and hierarchical models are partially degenerate. However, even in these cases the results for the monolithic model occupy a very narrow band of $I$($v_{2D}$) versus $t$ parameter space very close to zero, ($I$($v_{2D}$) $=$ -0.0066 $\pm$ 0.0036 even in the case of severe biases). If the observed $I$($v_{2D}$) of real star-forming regions deviates further than this from the expected value of Moran's $I$ statistic assuming no spatial autocorrelation, as defined in equation \ref{expected_val}, then it can be conclusively said that the data is inconsistent with the monolithic star formation model analogue presented in this paper, and this result can reasonably be extended to all formulations of that model without initial spatial-kinematic substructure that form in approximately stable state. As a result, calculating $I$($v_{2D}$) of a set of real star forming regions presents an exciting opportunity to provide strong quantitative evidence for/against the monolithic and hierarchical models of star formation. 

Examples of possible target regions include the Orion Nebula Cluster and the nearby young clusters in Taurus, such as NGC 1333. These regions are well-studied, meaning there is a wealth of data already available, and relatively close, meaning that the data available is of high enough quality that, per the findings of Sections \ref{impact_of_obs} and \ref{obs_bias_model_comp}, the results of this analysis would be robust.

\section*{Acknowledgements}

We would like to thank the referee for a kind and helpful referees report which improved the clarity and content of the paper. NJW and BA acknowledge a Leverhulme Trust Research Project Grant (RPG-2019-379), which supported this research. RJP acknowledges support from the Royal Society in the form of a Dorothy Hodgkin fellowship (grants DH150108 and CEC19\textbackslash100159).

\section*{Data Availability}

Access to the simulations this research is based upon are available by contacting B. Arnold at r.j.arnold@keele.ac.uk. The calculated $I$($v_{2D}$) and $I$($s$) as a function of time for the simulations are provided in online-only supplementary material.



\bibliographystyle{mnras}
\bibliography{complete_manuscript_file} 



\appendix 

\section{Generating substructured star forming regions} \label{frac_explanation}

Regions with substructured initial conditions are generated via a fractal algorithm which always produces regions with kinematic substructure, and which have a tunable amount of spatial substructure. In the substructured class, the spatial substructure is set to be high.

The algorithm to generate fractal initial conditions is now described. It is a minor variation on the algorithm explained and used in \citet{Goodwin04}, \citet{Allison10}, \citet{Arnold17}, \citet{DaffernPowell20} and \citet{BlaylockSquibbs22} among others. The original method generates fractal initial conditions by beginning with a cube of size unity. A single `parent' star is placed at the centre of this cube. The cube is then divided in two along each axis to form eight sub-cubes, and `child' stars are placed at the centre of these sub-cubes with some noise added to avoid an unrealistic regular structure. Next the parent star and a random subset of the child stars are deleted. The probability of deletion is the same for every child star and is abstracted from the user-defined fractal dimension, $D$. 

The surviving child stars then become parents themselves. Their host sub-cubes are further subdivided in two again along each axis, and new child stars are placed approximately in the centre of each; again, some noise is added to avoid an artificially grid-like structure. 

The parent stars and a random subset of these new children are deleted, and the process repeats until the desired number of stars for the initial conditions, $N$ has been overproduced. A sphere is then cut from the original box; stars outside this sphere are deleted. Random stars are then deleted until $N$ stars remain.

The degree of spatial substructure in the generated initial conditions is controlled by $D$. If $D$ is low few stars will survive each generation, leaving large areas of empty space where no more generations can propagate. Consequently, the resulting structure is highly spatially substructured. If, on the other hand, $D$ is large most child stars will survive, all areas of the fractal will be approximately equally populated and the resulting initial conditions will have low spatial substructure. 

Velocity substructure is engendered in the initial conditions as follows. The velocity of a given child star is assigned is: 

\begin{equation}
  \bm{v}_{child} = \bm{v}_{parent} + \xi\bm{\sigma}_{rand}
\end{equation}

\noindent where $\bm{{\sigma}_{rand}}$ is a random velocity vector and $\xi$ is a scaling factor that decreases with each generation. This results in more closely related stars having more similar velocities on average. Initial conditions generated by this method always have velocity substructure, this is independent of their degree of spatial substructure. 

As discussed in Section \ref{simulations}, stars are assigned masses drawn from the Maschberger IMF distribution \citep{Maschberger13}. Next the entire initial conditions are re-scaled to reflect the user's desired half-mass radius. Finally, the velocity vectors are scaled in order to set the virial ratio, $\alpha_{\rm{vir}}$, that is desired by the user. 

As previously mentioned, the method used to produce the initial conditions for the simulations in this paper is a minor variation on the widely used one described above. In the above method cubes are subdivided in two at each generation, and children are placed in the approximate centre of each sub-cube. This tends to leave a void in the centre of the parent box, which is particularly noticeable at the first generation. This results in a significant underdensity of stars in the centre of the completed initial conditions; they appear unrealistically hollow. In the method used in this paper, the boxes are divided in three along each axis at each generation, rather than in two, in order to prevent this. 

Additionally, rather than use a fractal dimension, $D$, to control the degree of spatial substructure we directly set the probability $P$ of child stars surviving to become parents. At the stage where child stars are culled prior to them becoming parents a value is drawn for each child star from a uniform distribution between zero and one. Those that have a value $< P$ drawn survive. There is an exception for this in the first two generations; it is required that at least six stars survive both of these generations to prevent an overly concentrated initial condition set being produced. If this requirement is not met the initial conditions are regenerated.

To generate the substructured initial conditions used in this paper $P$ is set to 0.2. While $P$ is used for convenience it is directly related to $D$. A $P$ of 0.2 translates to a $D$ of 1.535, which is a close proxy for the value $D = 1.6$ which has been used as a shorthand for high substructure in several numerical studies, e.g. \citep{Allison11, Parker14, Parker16b, Arnold17}. At this $P$ six generations of stars are required to produce sufficient stars for the spherical cutting stage. The exact diameter of the regions varies because while the half-mass radius is scaled to 2 pc in all cases the exact distribution of stellar masses is randomly drawn, so the width of the cut out sphere is not fixed. The mean diameter is 5.27 pc. This means that the size of the smallest cubes that contain children (i.e. the scale of the smallest structures in the region) is 5.27 pc divided $3^{g_n - 1}$ where the generation number, $g_n$ is six. Therefore the scale of the smallest structures is 0.022 pc.

In this paper, the noise added to the locations of child stars to avoid a grid-like structure takes the form of a displacement of the star from the centre of the sub-cube by a 3D vector. Each component of this vector is drawn from a Gaussian distribution of width $\iota$, defined as:

\begin{equation}
  \iota = \frac{1}{10L}
\end{equation}

\noindent and $L$ is the side length of the sub-cube. This width is chosen as the probability that a star will be be displaced outside of its host sub-cube is small (this would require a 5 standard deviation displacing vector), while still allowing for significant fluctuation from the rigid grid structure. 
For clarity, this random vector is redrawn for each star, rather than drawn once and applied to them all which would simply displace the entire generation.

In the implementation of the algorithm used in this paper $\bm{{\sigma}_{rand}}$ is a three dimensional vector where each component is randomly drawn from a Gaussian distribution centred on zero with width one. $\xi$ is defined such that

\begin{equation}
  \xi = \frac{1}{g_n}^{\frac{2}{3}}
\end{equation}

\noindent where $g_n$ is one for the first generation of stars, their children have $g_n = 2$, and so on). $\xi$ is chosen to be an inverse function of $g_n$, to ensure $\xi$ decreases with each generation, as described. This is taken to a power of one third in keeping with the decrease in spatial scale, and squared in order to provide a more realistic overall velocity dispersion.


\bsp	
\label{lastpage}
\end{document}